# Elastic Mid-Infrared Light Scattering: a Basis for Microscopy of Large-Scale Electrically Active Defects in Semiconducting Materials

## V. P. Kalinushkin, V. A. Yuryev[1] and O. V. Astafiev[2]

### General Physics Institute of the Russian Academy of Sciences
### 38, Vavilov Street, Moscow 117942, GSP-1, Russia

A method of the mid-IR-laser microscopy has been recently proposed for the investigation of the large-scale electrically and recombination active defects in semiconductors and non-destructive inspection of semiconductor materials and structures in the industries of microelectronics and photovoltaics. The basis for this development was laid with a wide cycle of the investigations on the low-angle mid-IR-light scattering in semiconductors. The essence of the technical idea was to apply the dark-field method for spatial filtering of the scattered light in the scanning mid-IR-laser microscope. This approach enabled the visualization of large-scale electrically active defects which are the regions enriched with ionized electrically active centers. The photoexcitation of excess carriers within a small volume located in the probe mid-IR-laser beam enabled the visualization of the large-scale recombination-active defects like those revealed in the optical or electron beam induced current methods. Both these methods of the scanning mid-IR-laser microscopy are now introduced in detail in the present paper as well as a summary of techniques used in the standard method of the low-angle mid-IR-light scattering itself. Besides the techniques for direct observations, methods for analyses of the defect composition associated with the mid-IR-laser microscopy are also discussed in the paper.

Special attention is paid upon potential applications of the above methods as characterization and testing techniques in the semiconductor science and industry. It is concluded that elastic mid-infrared laser light scattering is a basis for the development of a variety of research techniques and instruments which could be useful in different branches of basic and applied research work in the field of defect engineered semiconductors as well as for the development of devices for quality inspections in the semiconductor industry. Being contactless, non-destructive and non-polluting, techniques based on mid-infrared light scattering could also find many applications for automation of the technological process control as well as the process remote monitoring.

## I.    Introduction

A method of the low-angle mid-IR-laser light scattering (LALS)[3] has been successfully used for the investigation of semiconductors since the early 70-s. Initially, it was applied to study exiton drops at liquid helium temperatures but it was found before long that the intense scattering of light remained in Ge and Si even at room temperature when no exiton drops can exist[4] and even without photoexcitation[5]. It was likely the first direct observation of large-scale accumulations of electrically active (LSDAs) and recombination-active (LSRDs) defects in the bulk of semiconductor crystals.

Since that time, a number of works have been carried out in which LSDAs have been observed in such crystals as Si[5−7], Ge[8], GaAs[9−11] and InP[12−14] by means of LALS with angular resolution. Some new opportunities of LALS have been demonstrated and new methods have been developed on its basis.

It has been shown that the intensity of light scattering in LALS usually does not depend on the wavelength of the probe radiation at least in the wavelength interval of $\lambda \gtrsim 5$ μm. The theory predicts namely such behavior for the elastic scattering of light by free carrier accumulations (FCAs) whose dielectric functions are proportional to $\lambda^2$ unlike dielectric constants of different defects which do not depend on the incident light frequency (of course if there is no resonance). It was concluded that the scattering by FCAs dominated in this spectral interval, and the method for selection between FCAs and different scatterers was proposed on this basis which used CO and $CO_2$-lasers as sources of the probe radiation. [15,16]

When it was realized that the FCAs are formed because of the thermal ionization of some dissolved but aggregated impurities and defects (i.e. they are so called "impurity clouds" or other LSDAs), a method of the temperature dependent LALS was developed which enabled the estimation of the activation energies of the defects constituting LSDAs.[3,16−18] This method consists of measuring the temperature dependencies of the light scattering intensity at a fixed angle associated with some class of scatterers and the light scattering angular diagrams at different sample temperatures.

And at last, to investigate recombination properties of LSDAs and LSRDs (the latter are not necessarily LSDAs as well as the former often are not LSRDs) the technique of LALS with sample photoexcitation was developed.[3,4,19−22] Application of the light with quantum energy slightly less than the studied semiconductor band gap enabled the investigation of the defects in bulk crystals, whereas application of band-to-band photoexcitation enabled the exploration of the defects in subsurface layers of semiconductors.[21,22] This method obviously enables the detection of defects under coatings and near interfaces when appropriately choosing a pumping laser wavelength.

Amongst the physical results obtained by means of LALS with angular resolution, we would like to highlight the following ones.

The application of this method gave the evidence that the resolution of γ-ray detectors made of pure Ge was controlled by LSDAs— so called "oxygen clouds".[3,23]

The experiments on the effect of high temperature annealings, growth stoppage and variation of crystal growth rate on the light scattering by LSDAs enabled the determination of their formation temperatures, composition[24] and the origins of their formation in pure FZ Si and CZ Ge.[3,5,6,8,19,25−28] The experiments on doping with rapidly diffusing impurities yielded that the conductance type of the "impurity clouds" in pure p-Si and p-Ge is p-type one.[29,30,31] The estimates of activation energies for the centers constituting different LSDAs in n-Ge were made in Ref. 18 from the temperature dependent LALS.

Detailed research of CZ-grown Si targeted to classification of LSDAs was presented in Refs. 32–34. It was also shown in these works that thermal donors are likely contained in the LSDAs in CZ Si. In addition, it was demonstrated in Refs. 35,36 that a combination of the conventional and temperature dependent LALS techniques and the method of LALS with photoexcitation is very useful for the investigation of the internal gettering process. The formation the gettering defects in CZ Si:B as well as the impurity atmospheres around them was observed directly, and the thermal activation energies of point centers constituting the atmospheres after the gettering process were estimated for the samples obtained from different manufacturers and found to be different.

A series of works was devoted to detailed exploration of LSDAs in LEC InP and GaAs.[37–41] In these papers LSDAs were modeled by spherical p-type regions enriched with In or Ga microinclusions and $In_P$ and $Ga_{As}$ acceptors, respectively.

As it is seen from the above brief essay, sufficiently large work has been made by LALS to investigate such specific classes of defects as LSDAs in different semiconductors. We can say from our experience that LALS is one of the most (and likely the most) appropriate method for such kind of experiments. It is contactless, nondestructive, does not require complicated preparation of samples, and its sensitivity is very high (it can sense exhaustive free carrier concentrations in FCAs down to $10^{13}$ cm$^{-3}$). It might seem that one would not desire anything better, but unfortunately, LALS has two serious shortcomings which restrict the area of its possible applications and sometimes make interpretation of results rather ambiguous.

Firstly, using LALS one cannot estimate the concentrations of scatterers $C$ separately but only multiplied by the square of variation of the dielectric constant: $C/\delta\varepsilon|^2$.[2,3,33–41] To evaluate $C$ (and consequently $|\delta\varepsilon|$ which is necessary e.g. for estimating the activation energies of the centers constituting LSDAs[33–41]) one must use data of different techniques, such as EBIC, laser tomography, etching, *etc.*, that reduces the reliability of the information ob-tained. One can never vouch for the equivalence of defects registered in LALS measurements and those revealed by other methods.

Secondly, LALS does not enable the study of each defect alone and gives the information averaged over a group of defects with close parameters located in the probe beam. Moreover, if there exist several kinds of scatterers with close enough sizes, the interpretation of data becomes rather tricky.

The undoubted merit of LALS is, as mentioned above, its extremely high sensitivity, especially to FCAs.[3] LALS also often enables the determination of the distribution of point centers inside LSDAs.

The main objective of the present paper is to introduce two methods of laser scanning microscopy based on LALS. The first of these methods—referred to as the scanning mid-IR-laser microscopy or the scanning LALS (SLALS)—allows one to overcome the above disadvantages of the standard LALS. The second one—the optical beam induced mode of the scanning LALS (OLALS)—is an optical analog of EBIC and OBIC. It enables the investigation of LSRDs in near-surface regions of semiconductor wafers, under coatings, near interfaces, in bulk crystals and so on. But in contrast to EBIC and OBIC, it requires neither Schottky barrier nor p-n junction and any special preparation of wafer surfaces. Both methods are based on LALS physical principles and have all of its advantages.

The second aim of the paper is to consider possible practical applications of the latter methods as well as other LALS-based techniques for wafer and material testing and characterization in both microelectronics industry and research labs.

For the beginning we shall briefly remind the physics on which LALS is based and its main modes and opportunities. One can find some additional details in a large but rather old paper cited in Ref. 3.

## II.    Physical basis of LALS

### A.    Angular diagram of light scattering

There are many works in which the dependence of light scattering intensity on the observation angle is analyzed for scatterers with different optical densities, shapes and sizes. Rather complete theory of scattering is given e.g. in Refs. 42–44. Here we would like only to write some formulae which describe angular light-scattering diagrams of such specific objects as LSDAs.

Usually, LSDAs in modern crystals are "weak" optical inhomogeneities with small deviation of dielectric functions from those in crystal bulk outside LSDAs. So, they satisfy the Reyleigh-Gans condition:[42]

$$|\tilde{\varepsilon}| << \frac{\lambda}{2\pi a} \qquad (1)$$

where $\tilde{\varepsilon} = \delta\varepsilon/\varepsilon$ is a relative deviation of the dielectric constant in the defect, $\lambda$ is the scattered light wavelength inside the crystal, and $a$ is the linear size of inhomogeneity. In this case, the electric field of the light scattered by one inhomogeneity is:[44]

$$\vec{\mathbf{E}}_s = -e^{ikR} \frac{E_0^i}{4\pi R} \left[ \vec{\mathbf{k}}_s \left[ \vec{\mathbf{k}}_s \vec{\mathbf{G}} \right] \right] \qquad (2)$$

where the vector $\vec{\mathbf{G}}$ is given by the relation:

$$\mathbf{G}_j = \int_v \tilde{\varepsilon}_{jk} \, \mathbf{e}_k^i \, \exp\left(i\vec{\boldsymbol{\kappa}}\vec{\mathbf{r}}\right) d^3 r \,, \qquad (3)$$

and the unit vector $\mathbf{e}_k^i$ in the direction of the incident field $\mathbf{E}_k^i$ is determined by the equation:

$$\mathbf{E}_k^i = E_0^i \, \mathbf{e}_k^i \, e^{i\vec{\mathbf{k}}\vec{\mathbf{r}}}. \qquad (4)$$

The following designations are used in the equations ( 2 )–( 4 ): $E_0^i$ is the amplitude of the incident wave; $\vec{\mathbf{k}}$ is its wave-vector; $\vec{\mathbf{k}}_s$ is the wave-vector of the scattered light, both inside the crystal; $\vec{\mathbf{R}}$ is the radius-vector of the observation point; $\vec{\mathbf{r}}$ is the radius-vector inside the inhomogeneity; $\vec{\kappa} \equiv \vec{\mathbf{k}}_s - \vec{\mathbf{k}}$; the scattering is elastic: $k_s = k$. The integration is made over the volume of the inhomogeneity $v$.

It is easy to obtain for identical isotropic spherically symmetrical inhomogeneities with mean concentration $C$ and the dielectric functions

$$\tilde{\varepsilon}(r) \equiv \tilde{\varepsilon}_m \phi(r,a) \tag{5}$$

that the light intensity divided by the probe beam power $W$ is:

$$\frac{I}{W} = L C \left| \tilde{\varepsilon}_m \right|^2 a^6 k^4 \cos^2 \theta \cdot \left| \Psi(\kappa,a) \right|^2, \tag{6}$$

$$\Psi(\kappa,a) = \int_0^\infty \phi(\xi) \frac{\sin(\kappa a \xi)}{\kappa a \xi} \xi^2 \, d\xi \tag{7}$$

where $\xi = r/a$, $\phi(r,a) \equiv \phi(\xi)$ is the profile function of the dielectric constant distribution inside the scatterers, $\theta$ is the scattering angle inside the crystal, $\tilde{\varepsilon}_m$ is the maximum value of $\tilde{\varepsilon}$, $\kappa \approx k\theta$ for the small angles, and $L$ is the sample thickness.[3] A Poisson distribution of scatterers in the crystal is assumed. The averaging is made over some interval of angles $\delta\theta$ ($\lambda/a >> \delta\theta >> \lambda/D$, $D$ is the probe beam diameter) for each value of $\theta$.[45]

It is seen from ( 6 ) and ( 7 ) that by measuring the light scattering diagram one can determine the profile function $\phi(r,a)$, the sizes of scatterers $a$ and the value of $C \left| \tilde{\varepsilon}_m \right|^2$.[46]

If the LSDAs are not spherical (as well as in the case of anisotropy) the equations can be derived from ( 2 ), ( 3 ) to replace ( 6 ), ( 7 ). In this case one must measure also the dependence of light scattering diagrams on the sample orientation with respect to the probe beam polarization and/or measurement plane to obtain information on the shape of the scattering defects.[47] A schematic diagram of the instrument for measuring the angular diagrams of LALS is shown in Fig. 1. Let us illustrate the above with an example. Typical light-scattering diagrams are presented in Fig. 2. Two main types of defects are revealed in CZ Si:B crystals (Fig. 2 (a)): cylindrical defects with the diameters from 3–4 to 8–10 μm and lengths from 15 to 40 μm ($\theta < 5°$) and spherical ones with the sizes from 5–8 to 20 μm $(\theta > 5°)$.[33,34] The concentration $C$ of the spherical defects grows by about two orders of magnitude as a result of the internal gettering process exposure whereas the concentration of the cylindrical ones remains unchanged (Fig. 2 $(b)$).[35,36]

## B.  Dielectric function of LSDA

As the LSDAs registered by means of LALS are the regions enriched with ionized impurities and/or intrinsic defects (see the bibliography cited in the Introduction), they are the accumulations of free carriers (FCAs). LALS uses light with a wavelength $\lambda$ of 5–10 μm as the probe radiation which makes it sensitive to FCAs (namely due to this circumstance LALS may be used efficiently to study LSDAs). Let us write down the dielectric function for FCAs $\widetilde{\varepsilon}(r)$. Since the light frequency as a rule satisfies the condition $\omega >> \tau^{-1}$ ($\tau$ is the relaxation time for the free carrier), the dielectric function is given by the equation: [48]

$$\widetilde{\varepsilon}(r) = -\frac{4\pi e^2}{\varepsilon \omega^2}\left(\frac{n(r)}{m_1^*} - \frac{n_\infty}{m_2^*}\right) \tag{8}$$

where $n(r)$ is the distribution of free carrier concentration inside FCA, $n_\infty$ is the free carrier concentration outside FCA, $m_1^*$ and $m_2^*$ are the ohmic effective masses of current carriers inside and outside FCA respectively, $\omega$ is the incident light cyclic frequency, $e$ is the electron charge.

It is seen from ( 8 ) that $\widetilde{\varepsilon}(r) \sim \lambda^2$ whereas for electrically inactive defects, $\widetilde{\varepsilon}(r)$ is independent of $\lambda$. Substituting ( 8 ) in ( 6 ), ( 7 ) one can easily be convinced that the light scattering intensity for all inhomogeneities except for FCAs drops rapidly with the growth of $\lambda$ whereas the scattered light intensity for FCAs is independent of $\lambda$ or in some cases even grows (the method described in Refs. 15,16 for determination of the nature of scatterers is based namely on this conclusion). As a result, when $\lambda$ is large enough (e.g. as large as 5–10 μm) the scattering by FCAs starts dominating the scattering by other defects, of course, for sufficiently perfect technological crystals such as modern Ge, Si, GaAs and InP.

The scattering diagrams for CZ Si measured at the wavelength $\lambda$ of 10.6 µm, 5.4 µm and 1.15 µm are given in Fig. 3. For 5.4 µm and 10.6 µm, the curves are well fitted, whereas for 1.15 µm the scattering is remarkably greater due to small submicron defects. This proves that the scattering by FCAs is obsereved at the wavelengths of 5.4 µm and 10.6 µm and the impact of defects with $\tilde{\varepsilon}$ independent of $\lambda$ (e.g. precipitates, *etc.*) is dominating at 1.15 µm.

## C. Dependence of light scattering on sample temperature.

It is obvious from the above that the measurement of the temperature dependence of the light scattering is in a sense analogous to the study of the temperature dependence of the material conductivity: both methods record the free carrier concentration as a function of the sample temperature, but in the second case the carriers are distributed in the whole crystal bulk whereas in the first case they are located in FCAs (i.e. in LSDAs). Hence to determine the thermal activation energies of the centers constituting LSDAs, the dependence of the light-scattering intensity on the sample temperature should be measured.

Usually, these dependencies are steps of rather small height (Fig. 4).[13,14,16,33–41] Refs. 37,38 give the formulae for estimating energy level location from the LALS temperature dependencies:

$$E_i = E_F - k_B T_0 \ln(g) \tag{9}$$

where the Fermi energy $E_F$ is calculated from the carrier concentration (hole concentration for p-type FCAs, electron for n-type ones)

$$n = N_{v,c} \exp\left( -\frac{\left| E_F - E_{v,c} \right|}{k_B T_0} \right) \tag{10}$$

at the temperature $T_0$ corresponding to the scattering intensity

$$I = \frac{1}{2}\left( \frac{I_1 + I_2}{2} + \sqrt{I_1 I_2} \right). \tag{11}$$

Here $I_{1,2}$ are the scattering intensities corresponding to the upper and lower edges of the step respectively, $N_{v,c}$ is the density of states in the valence ($v$) or conduction ($c$) band, $E_{v,c}$ is the top of the valence band ($v$) or the bottom of the conduction band ($c$), the subscripts $c$ or $v$ are used for n- or p-type FCAs respectively, $k_B$ is the Boltzmann constant, and $g$ is the degeneracy ratio for the level whose energy is estimated. The value of $n$ is calculated from experimental light scattering diagrams using equations ( 6 )–( 8 ).

Fig. 4 presents temperature dependencies of the light scattering intensity extrapolated to zero scattering angle for different LSDAs in different semiconductors. The activation energies are evaluated with respect to the corresponding band (conduction or valence) depending on the conductance type assumed or determined for each kind of defects.[16,18,37–41] Two steps observed in Ge correspond to two different point centers which freeze out at different sample temperatures, have different thermal activation energies, but contained in the same LSDAs.

Temperature dependencies of the light scattering intensity depicted in Fig. 5 (*a*) demonstrate the behavior of point centers in two main classes of defects in CZ Si:B, and the scattering diagrams plotted in Fig. 5 (*b*) show the moment when the spherical LSDAs have already been frozen out whereas the cylindrical ones have stayed as activated as they were at room temperature.[32–34,36] It should be emphasized that the spherical defects play an important role in the internal gettering process in CZ Si:B (Fig. 2 (*b*)), while the cylindrical ones dominate in the as-grown material.[35,36]

Now we would like to remark that applying the above methods one can calculate the profile function $\phi(r,a)$ for the scattering defects using experimental scattering diagrams and formulas ( 6 )–( 8 ) and the distribution of $n(r)$ inside them, and estimate the activation energies of the centers composing the defects from the light scattering temperature dependencies and expressions ( 9 )–( 11 ); then the distribution of the defect-forming point centers inside LSDAs can be easily obtained as well as the spatial potential relief caused with the investigated defects (Fig. 6).

**D.    LALS with sample photoexcitation**

As it is seen from the formulae ( 6 )–( 8 ), LALS is sensitive to all non-uniformities of the free carrier distribution in crystals if they have an appropriate size and carrier distribution inside, and no matter whether they are inherent to the crystal itself or induced artificially from outside during measurements e.g. by exciting excess carriers. So, if one has a willingness to investigate the recombination properties of LSDAs or to reveal some defects which do not give rise to extensive enough atmospheres of ionized centers and, as a consequence, to FCAs, but interact with non-equilibrium carriers—so called large-scale recombination-active defects (LSRDs) or large-scale gluing centers (LSGCs)[3]—one could use optical excitation of non-equilibrium carriers in studied samples with a wide laser beam.

If light with an energy slightly less than the sample bandgap is used for photoexcitation, sufficiently uniform distribution of excited electron-hole pairs is generated in the crystal bulk everywhere except for the vicinities of LSRDs and LSGCs. The deviation of dielectric function arises near LSRDs and LSGCs which can be registered by means of LALS. This method was actively used e.g. for investigation of the defects in pure FZ Si and Ge. Many references on the works made by this method are cited in the Introduction of the current paper.

If light with an energy greater than the sample bandgap is used, one can investigate LSRDs and LSGCs directly in near-surface regions of crystals which are often of primary interest for the purposes of microelectronics. This technique was demonstrated for the first time in the Refs. 21,22 for the crystals of Ge taken as an example.

An appropriate choice of the exciting light source and conditions of measurements (the pumping light wavelength $\lambda_{ex}$, pulse duration and frequency, measurement gate, *etc.*) will enable the investigation of epitaxial layers and surfaces under them, surfaces under dielectric coatings (e.g. Si coated with $SiO_2$ film), interfaces of semiconductors with semiconductors or dielectrics and so on.

The experimental setup for this method is practically the same as that for ordinary LALS. The only addition is the laser for photoexcitation (Fig. 1).

Fig. 7 (*a*) shows the light-scattering diagrams for mechanically and chemical-mechanically polished samples of pure single-crystalline Ge obtained with and without band-to-band (surface) photoexcitation ($\lambda = 10.6$ μm, $\lambda_{ex} = 1.06$ μm). The effect of the damaged layer is clearly seen for the mechanically polished samples: the intense "plateau" in the diagrams is attributed to the impact of the light scattering by inhomogeneities of non-equilibrium carrier distribution caused by small-sized LSRDs located near the surface. For the chemical-mechanically polished samples, the recombination activity of LSDAs in Ge shown for the first time in Ref. 19 using LALS with bulk excitation is manifested in the diagrams: they are similar for the measurements with and without excitation. The diagrams co-incide for both samples without excitation. Fig. 7 (*b*) demonstrates the strongly super-linear dependency of the light scattering intensity on the exciting light power which confirms the assumption that non-uniformities in the non-equilibrium carrier distribution around LSRDs is responsible for the scattering of light observed in this experiment. This cubic dependency is characteristic for the mechanically lapped samples[21,22,49–51] as distinct from the chemical-mechanically polished ones as well as the bulk semiconductors[1] investigated by LALS with bulk excitation which always show the routine second-power dependency[49–51] corresponding to the linear recombination.

The internal gettering process can be efficiently investigated by means of LALS with photoexcitation too. Light scattering diagrams measured under quasi-bulk excitation for a wafer of CZ Si:B before and after the gettering treatment are plotted in Fig. 8. The formation of small recombination-active defects in the crystal bulk is clearly seen. These defects seem to be responsible for the gettering property of the crystal interior and the formation of the impurity atmospheres observed in Fig. 2 (*b*).

## E. Discussion of LALS

It is seen from the above that LALS gives unique opportunities for investigations of crystal imperfections and non-uniformities and effect of technological treatments on them, as well as the study of process induced defects. Its sensitivity is very high due to the application of the Mach–Zehnder interferometer as a laser heterodyne:[3,52] it allows one to register FCAs with free carrier concentration down to $10^{13}$–$10^{14}$ cm$^{-3}$. It has a high resolution when evaluating the sizes of defects. The effective sizes of domains with diameters down to 1–2 μm can be still measured although the wavelength of probe radiation is 5–10 μm (the upper size detection threshold of 30–40 μm is controlled by the probe IR-laser beam width). It enables the investigation of the composition of defects in the contrast to e.g. EBIC, OBIC, etching or laser tomography. It allows one to study wafers without any special preparation except for polishing on both sides. And lastly, it can be modified to study recombination-active defects, just like OBIC and EBIC but without Schottky barrier or p-n junction, special etching of the studied surface and other specialized preparation procedures which the latter techniques require,[53] and even under coatings and near interfaces. It appeared to be among the most informative techniques for the investigations of such important processes as the internal gettering one. In addition, it enables the *in situ* process investigations, e.g. if the studied sample is placed in a specialized furnace, *etc.*, or even remote monitoring of processes in a technological cycle.

Unfortunately, as already mentioned in the Introduction, LALS has two main closely connected shortcomings. First of all, it cannot yield the values of scatterers concentrations $C$, which are necessary for obtaining precise estimates of the free carrier concentrations inside LSDAs and, hence, the values of the activation energies of the centers constituting them. The second demerit of LALS is its inability to discriminate between separate scatterers, so it yields only average values of estimated parameters for each group of defects with close dimensions.

To overcome these shortcomings, a method for visualization of FCAs (scanning LALS or SLALS) has been developed.[41,49,54 −58] This method is based on the same physical principles as LALS, hence it possess all its unique opportunities, including the ability to study LSRDs which has found its most promising realization in the optical beam induced LALS (OLALS) mode of SLALS.[49−51,58,59]

## III.    SCANNING MID-IR-LASER MICROSCOPY (SCANNING LALS)

### A.    The basic instrument

The main distinction of the scanning LALS (SLALS) from the classical LALS consist in the following: a defect image is formed in scattered light in SLALS, whereas the information about defects is obtained from the scattering diagrams without image formation in LALS. A functional diagram of an ideal arrangement for SLALS is shown in Fig. 9. A parallel-sided semiconductor crystal polished on both sides (e.g. a standard technological wafer) is illuminated with the plane wave of the mid-IR-laser emission. The sample is located in the front focal plane of the lens L1. Let a defect be in the crystal volume in the front focus of L1. It scatters the probe wave, producing an additional wave which diverges into the diffraction angle. The lens L1 condenses the plane probe wave in the back focal plane—which coincides with the front focal plane of the lens L2—where a small round opaque screen (or a mirror turned to some angle to the focal plane) with the diameter $b_1$ and an aperture with the diameter $b_0$ are placed on the optical axis to remove the probe wave radiation. The scattered wave, being practically a plane wave after L1, passes to the lens L2 almost without losses—if $b_1$ is small enough and $b_0$ is large enough—and is condensed in the back focus of L2 where an IR photodetector is located. ($f_1, f_2, D_1$ and $D_2$ are the focal distances and apertures of the lenses L1 and L2, respectively.) Scanning the sample, the images of defects can be obtained in the scattered rays.

This scheme was proposed for SLALS for the first time in Refs. 54,55. It represents the well-known method in microscopy of the dark field[60−63] (which was formerly referred to as the central dark ground[62] or Töpler-Foucault[60,61] method, it was especially actively used in the optical microscopy for revealing phase objects until the phase contrast method[61−70] was developed by Zernike[71]) and comprises a space-frequency filter (often used in holography) which suppresses the light components with low space frequencies (e.g. the probe radiation transmitted through the crystal and, alas, the images of too large objects) and the harmonics with too high space frequencies (which are limited with the microscope aperture).[56] The images of objects with the characteristic radii of

$$\frac{2j_{1,1}f_1}{kb_0} < a < \frac{2j_{1,1}f_1}{kb_1} \qquad (12)$$

are transferred through the optical system almost without distortions and light losses ($j_{1,1}$ is the first zero of the Bessel function of the first kind of order unity). The images of larger objects are suppressed, while the images of minor ones are blurred into spots with the diameters of about $4j_{1,1}f_1/kb_0$.

Let us dwell on the analysis of the SLALS-microscope in some more details. Like in Sec. II A, we assume the dielectric function of a defect to be different from that of the crystal outside the defect by the value of $\delta\varepsilon(x,y,z)$. There are two main classes of defects common for the technological semiconductors: i) phase objects like e.g. FCAs, and ii) defects absorbing the incident light like e.g. some foreign inclusions. In general, most of defects can be considered as a superposition of these cases.

Keeping in mind ( 1 ), the total field after a phase object is:

$$\vec{E}_\Sigma = \vec{E}_0 \exp\left[-i\varphi(x,y)\right] \approx \vec{E}_0 - i\varphi(x,y)\vec{E}_0 = \vec{E}_0 + \vec{E}_1(x,y) \qquad (13)$$

where the phase shift on the defect $\varphi(x,y) \approx ka\,\widetilde{\delta\varepsilon}(x,y)/2$, $(x,y)$ are the coordinates in the front focal plane of L1. Hence, the total field after the defect may be considered as a superposition of two waves: an undisturbed incident (probe) wave $\vec{E}_0$ and a wave arising on the de-fect $\vec{\mathbf{E}}_1(x,y) = -i\varphi(x,y)\vec{\mathbf{E}}_0$ whose amplitude is proportional to the phase shift on the defect (and therefore, $\tilde{\varepsilon}(x,y)$ or, following ( 8 ), $n(x,y)$) and phase is shifted with respect to the phase of $\vec{\mathbf{E}}_0$ by the value of $-\pi/2$.

Analogously, we have for an absorbing defect:

$$\vec{\mathbf{E}}_\Sigma \approx \vec{\mathbf{E}}_0 - \alpha(x,y)\vec{\mathbf{E}}_0 = \vec{\mathbf{E}}_0 + \vec{\mathbf{E}}_1(x,y) \tag{14}$$

where $\alpha(x,y)$ describes the absorption; the phase of the field $\vec{\mathbf{E}}_1(x,y)$ is shifted with respect to the phase of $\vec{\mathbf{E}}_0$ by the value of $\pi$.

In both cases, an additional wave $\vec{\mathbf{E}}_1(x,y)$ of limited lateral dimensions (both $\varphi(x,y)$ and $\alpha(x,y)$ equals zero everywhere except for the defect region) arises at the defect and then diverges into the diffraction angle $\beta > \lambda/a$. It is described more accurately with equations ( 2 ) and ( 3 ).

Let us study the limitations imposed by the microscope on the sizes of defects that the images can pass through the optical system without distortions or losses of light, and the efficiency of the suppression of the interfering probe radiation on a detector. If the field propagating from the front focal plane of L1 is given by the equation:

$$\vec{\mathbf{E}}(x,y) = \vec{\mathbf{E}}_0\,\Phi(x,y)\exp\left(-i\vec{\mathbf{k}}\vec{\mathbf{r}}\right), \tag{15}$$

then one can obtain the following expression for the field in the back focal plane of L2:

$$\vec{\mathbf{E}}''(x'',y'') = -\vec{\mathbf{E}}_0 \frac{f_1}{2\pi f_2} \int_{-\infty}^{+\infty}\int_{-\infty}^{+\infty} \Phi(\xi,\eta)\Lambda\left[\sqrt{(\xi+p)^2+(\eta+p)^2},b_1,b_0\right]\mathrm{d}\,\xi\,\mathrm{d}\,\eta \tag{16}$$

where $(x'',y'')$ are the coordinates in the back focal plane of L$_2$, $p = x''f_1/f_2$, $q = y''f_1/f_2$, and

$$\Lambda\left[r,b_1,b_0\right] = \frac{b_0 k\,\mathrm{J}_1\!\left(\dfrac{b_0 k r}{2 f_1}\right)}{2 f_1 r} - \frac{b_1 k\,\mathrm{J}_1\!\left(\dfrac{b_1 k r}{2 f_1}\right)}{2 f_1 r} \tag{17}$$

is the point spread function of the dark-field microscope; $J_1(\zeta)$ is the Bessel function of the first kind of order unity. (See also Note 72.)

It is easy to obtain from ( 16 ) and ( 17 ) that the field in the back focal plane of L2 for defects satisfying the condition ( 12 )

$$\vec{\mathbf{E}}''(x'', y'') \approx -\vec{\mathbf{E}}_0 \frac{f_1}{f_2} \Phi(-\frac{f_1}{f_2} x'', -\frac{f_1}{f_2} y'') \qquad (18)$$

which demonstrates that the SLALS microscope, and like the conventional LALS, enables the investigation of the profile functions $\phi(r)$ of the defects and the evaluation of all the parameters of the defects discussed in Sec. II.

Then one can obtain from ( 16 ) and ( 17 ) for the beam with axial symmetry:

$$\vec{\mathbf{E}}''(0) = -\vec{\mathbf{E}}_0 \frac{f_1}{f_2} \int_0^{+\infty} \frac{d \Phi(r)}{d r} \left[ J_0\left(\frac{b_0 k r}{2 f_1}\right) - J_0\left(\frac{b k r}{2 f_1}\right) \right] d r \qquad (19)$$

where $J_0(\zeta)$ is the Bessel function of the first kind of order zero.

It is seen from ( 19 ) that to efficiently suppress the field of the probe beam on the receiver it is necessary to have its profile function $\Phi(r)$ smooth enough and the distance on which the derivative $d\Phi(r)/dr$ drops to zero be much greater than the periods of the Bessel functions in ( 19 ). In other words, high-frequency harmonics of $\Phi(r)$ for the probe beam must be minimized (diffraction effects on the beam edges and so on).

For the Gaussian field distribution in the probe beam or in the wave $\vec{\mathbf{E}}_1(x, y)$ arising due to a defect—the latter takes place if the defect has the Gaussian dielectric function profile $\phi(r)$ [45]—

$$\Phi(r) = \exp\left\{ -2\left(\frac{r}{a_b}\right)^2 \right\} \qquad (20)$$

where $a_b$ is the beam diameter which satisfies the condition $a_b^{-1} << k b_1 / 4\sqrt{2} f_1 << k b_0 / 4\sqrt{2} f_1$, the field distribution on the receiver is:

$$\vec{E}''(0) \approx -\vec{E}_0 \frac{f_1}{f_2} \exp\left\{-\frac{1}{2}\left(\frac{b_1 k a_b}{4 f_1}\right)^2\right\}. \qquad (21)$$

The field $E''(0)$ drops rapidly with the growth of the shading screen size $b_1$. For the screen diameter of only

$$b_1 = 6\sqrt{2}\left(\frac{2 f_1}{k a_b}\right) \qquad (22)$$

the field $E''(0)$ is suppressed by about 8000 times. This means that the intensity is reduced by about $6\times10^7$ times. Thus, in the case of Gaussian probe beam, it is sufficient to cover a few first Fresnel zones with the shading screen to eliminate the influence of the probe beam. Moreover, it appears from ( 21 ) for the screen from ( 22 ) that even defects which scatter only $2\times10^{-8}$ of the incident radiation power giving rise to the wave $\vec{E}_1(x, y)$ described with expressions ( 15 ),( 20 ) which has the diameter $a_b$ as large as one sixth of the probe beam diameter are still discerned through the system almost without suppression of their images intensity and dramatic reduction of contrast.

The sensitivity of SLALS in the present realization is high but somewhat lower than that of LALS (although it is not much lower even now: the minimum free carrier concentration in FCAs, which is senced by the used laboratory prototype of the microscope, is evaluated as $\Delta n_{\min} \approx (1-5)\times10^{15}\,\mathrm{cm}^{-3}$ and might be greatly reduced—down to $10^{13}\,\mathrm{cm}^{-3}$—by improvement of the instrument). The LALS setup represents the laser heterodyne (the Mach-Zehnder inter-ferometer);[3,52] in addition, LALS registers light from a great number of scatterers. Both these factors give LALS its exclusively high sensitivity in spite of the small aperture controlled with the detector size (Fig. 1). SLALS sensitivity is given only by its relatively large aper-ture, although the heterodyne device like that described in Ref. 73 is now under the develop-ment. The latter scheme will likely enable the development of the tomographic microscope with the longitudinal selectivity of about $2\lambda/\aleph^2$ ($\aleph$ is the numerical aperture),[52] and then the sensitivity of the heterodyne SLALS-microscope will be even higher than that of LALS.

## B.    Optical beam induced scattering mode (OLALS)

This mode of SLALS was proposed by us in Refs. 49–51,58,59,74 as a scanning modification of LALS with sample photoexcitation[3,4,19–23] (Fig. 9). The method, as well as LALS, can work in two regimes: bulk or surface photoexcitation. The regimes are different only by the choice of pumping laser: the first regime requires a laser with quantum energy less than the sample bandgap,[75] whereas the second one uses a laser with quantum energy greater than the bandgap; in general, both regimes are quite analogous.

Let us dwell on the surface excitation. The essence of the method consists in the following. A highly focused beam (unlike LALS with photoexcitation where a wide beam is used) generates excess carriers in the sample, in the chosen case in its near-surface region. If the "droplet" of the generated carrier (we mean the crystal domain where the concentration of the non-equilibrium carrier is high enough and the rim of its distribution is sharp enough) is as small as it is required in ( 12 ) the scattered mid-IR-laser light of the SLALS microscope reaches the photodetector. Its intensity is proportional to the square of the generated carrier concentration in the spot. The sizes of the "droplet" are controlled by both sizes of the exciting laser spot, the surface recombination velocity and the carrier lifetime and diffusivity. But even when the diffusion length is high (e.g. in Si with ~ 10-µs lifetime), a considerable inhomogeneity having rather rapid lateral decay in the sub-surface region remains in the carrier distribution due to the surface recombination.[78,79,80] This inhomogeneity is sufficient to be sensed by the SLALS microscope. The carrier concentration in the "droplet" is controlled with the non-equilibrium carrier effective lifetime in a given place on the sample, so we can speak about the recombination-contrast or effective lifetime imaging of semiconductors. In this sense, the closest to OLALS method was proposed in Ref. 81 in which free-carrier absorption was applied. Similar optical beam induced absorption sub-mode (bright field) is available in OLALS too, but the contrast is higher in scattered rays. [82]

The microscope resolution in OLALS is controlled by the effective dimensions of the crystal domain, in which the non-equilibrium carriers exist. In order to reach the maximum resolution, the exciting beam must be focused to such a spot size that will not affect the effective sizes of the domain. Thus the latter would be governed only by the carrier diffusion coefficient, surface recombination velocity and the carrier lifetime. So, the resolution upper limit of OLALS depends only on the studied sample properties rather than the diffraction limit of the microscope optical system.

It is clear that OLALS is a very close analog of the electron or optical beam induced current (EBIC or OBIC), but in distinction with the latter ones, OLALS requires neither a Schottky barrier nor a p–n junction. It also does not require any special preparation of surfaces. But the most important property of the developed method is its ability to give an information from interfaces and surfaces covered with coatings and epilayers, until metallized.[51,58]

Let us illustrate the above with three contrasting examples.

Fig. 10 demonstrates corresponding pairs of SLALS and OLALS micrographs for single-crystalline CZ Si:P wafers ($\rho = 4.5\ \Omega\ cm$) from the CCD manufacturing cycle. The images of LSDAs situated in the bulk of the substrates are seen in all the SLALS pictures as white spots (the whiter the image, the higher the free carrier concentration is). LSRDs located in sub-surface (or sub-interface) layers of the substrates are manifested in the OLALS pictures as dark areas (the darker the image, the lower is the excess-carrier concentration—and also shorter lifetime). The latter defects are distributed along the crystallographic directions and seem to be ones responsible for video defects in CCDs. LSDAs may also give rise to video defects, especially if located in the vicinity of the working layers, although their effect on the video signal of CCDs is doubtful if they are situated deep in the crystal volume.

SLALS and OLALS micrographs for single-crystalline $Si_{1-x}Ge_x$ alloy with Ge content from 2.2 to 4.7 at. % —a promising material for solar cells—are presented in Fig. 11. Two areas were revealed in the X-ray patterns of these crystals: the area free of striation and dislocations around the wafer centers (area I) and the area containing striation and dislocations in the periphery of the wafers (area II).[74] SLALS pictures shows the striation in the area II and no striation in the area I. OLALS pictures demonstrate that no or low (Fig. 11 (h)) recombination contrast is usually caused with the grown-in striations in the crystal bulk, although high contrast was revealed in Fig. 11 (l). The second type of defects manifested as dark stripes in the OLALS micrographs (Fig. 11 (b),(l)) can likely be identified as dislocations and dislocation walls which are registered in X-ray patterns of the area II and revealed by etching. The last type of defects observed are those seen as black spots in the OLALS patterns (Fig. 11 (b),(d),(f),(h),(j)). They are present in both areas and have a non-dislocation origin. Some non-dislocation defects were found in both areas by the selective etching which may be similar to those revealed by OLALS. The latter defects seem to be the main lifetime (and cell efficiency) killing extended defects in the studied material.

OLALS micrographs of multicrystalline Si for solar cells obtained in the induced absorption sub-mode are depicted in Fig. 12. Dark images of grain boundaries are clearly seen in both pictures. The images are formed due to the lower non-equilibrium carrier concentrations generated by the focused pumping beam in the vicinities of the strongly recombination-active grain boundaries in comparison with those generated when the probe beam scans the sample far from the grain boundaries. As a consequence, a lower concentration of the electron–hole pairs diffuses in the crystal bulk where the probe mid-IR light is absorbed by the free carriers. Fig. 12 shows the applicability of OLALS for investigations and inspections of the grain boundary passivation efficiency.

Three dependencies of the detector signal on the exciting beam power are plotted in Fig. 13. Square dependence was obtained for the chemical-mechanically lapped surface of a Si wafer which corresponds to the linear recombination. Like formerly determined for Ge crystals (Fig. 7 (b)), the third-power dependence was obtained for the mechanically polished surface. The explanation of this behavior may be as follows: perhaps, the recombination is linear too in this case and the carrier scattering on charged linear defects in the damaged sub-surface layer results in this unusual dependence (see note 83 for details). For the induced absorption, the linear dependence was obtained for both mechanically and chemical-mechanically polished sides (the line for the former case is not shown in the graph because it goes parallel to the line for the latter case, only the values are about 10 times less), which confirms that the IR absorption takes place mainly in the crystal bulk where the excited carrier diffuses.

Modifying the above instrument, it is also possible to create a kind of tomographic microscope on the basis of OLALS, and this problem does not seem to be very difficult. In addition, a combined instrument with the photoluminescence mapping (PL mapping) would be an even more powerful tool, especially if a spectrometer would be attached to the PL detector. In such an instrument, the photoexcitation for OLALS and PL might be given with the same focused laser beam, so the simultaneous PL and lifetime imaging with the local spectral analysis and radiative and effective lifetimes measurements would be possible. It is clear that the direct optical analog of the low-temperature EBIC [84,85] is easy to be developed on the basis of OLALS by placing a sample in an optical cryostat as is done in the temperature dependent LALS. The temperature-dependent SLALS of separate defects together with the defect profile function of the basic SLALS-instrument would give a combined measurement device with enhanced characterization abilities. In addition to PL and lifetime mapping, simultaneous sample mapping at $\lambda \sim 1$ μm would reveal the same defects as the laser tomography does.[86,87] So, combining all these techniques in one device would give a unique instrument for comprehensive defect investigations in semiconductors.

### C. Laboratory and industrial applications

Let us dwell on some possible specific applications of the techniques based on LALS for laboratory investigations and as the checking technique in the industrial microelectronics, optoelectronics and photovoltaics. The conventional LALS and its modifications seem to be more suitable for the development of industrial testing devices, including e.g. automated in-line or remote testing and process monitoring systems. We shall not consider such applications here as well as technical approaches to them. An emphasis will be made on the techniques based on SLALS which seem to be of primary interest for more complicated laboratory inspections and research work.

1. *Inspection of semiconductor wafer homogeneity.* As mentioned above, SLALS enables the observation of LSDAs with the sizes from several μm to several tens μm with point detect concentration in each of them down to $10^{13}$ cm$^{-3}$. This method permits the investigation of the LSDA composition and the influence of thermal treatments and operations of an industrial technological cycles on them. Wafer mapping is possible by means of SLALS even under coatings and layers. Incoming inspection and technological step checking with posterior utilization of substrates in the production process are also possible. The disadvantage of SLALS (as well as LALS) is its inability to discriminate between LSDAs situated in crystal bulk and ones located in near-surface layer. To remove this demerit, SLALS tomography with longitudinal selectivity down to 10–20 μm is now under developed. It will enable the testing of "working" near-surface layers of wafers and epitaxial layers for LSDAs.

2. *Inspection of presence of LSRDs in the bulk, near-surface, near-interface and epitaxial layers.* Methods of OLALS may be used for such studies. Like in the above case, a significant advantage of the technique is its ability to test multilayer structures including layers covered with different layers. In addition, there are no limitations on sample size and resistivity in OLALS. So wafer mapping, incoming and step testing with subsequent utilization of substrates in production cycle are possible.

Testing for LSRDs in near surface and near-interface regions of semiconductor wafers and inspections of quality of surface or interface preparation (e.g. the quality of surface polishing or washing) is one of the possible applications of OLALS. We investigated the quality of wafer polishing for Si wafers produced by different manufacturers and subjected to different polishing procedures and always revealed an inhomogeneity of non-equilibrium carrier lifetime distribution which was likely conditioned with the presence of polishing-induced subsurface defects. We also revealed such non-uniformity in the vicinity of Si–SiO$_2$ interface in the wafers subjected to the oxidation procedure.

Inspection for LSRDs in the substrate volume, including tomography, is also possible, but this is likely of interest for the production of γ-ray detectors, nuclear-ray counters, volume photodetectors, *etc*. Fruitfulness of the application of OLALS for the inspection of the gettering processes will be emphasized below.

3. *Testing of doped areas.* SLALS may be also used for the inspection of such parameters of doped domains of semiconductor structures as their sizes, concentration of free carrier in them and surface resistance. The inspection of these parameters is also possible even after different layers are grown and coatings are given (until metallized). The domains with sizes greater than 1 μm and variation of free carrier concentration greater than $10^{13}$ cm$^{-3}$ might be tested.

4. *Inspection of gettering process efficiency.* We would like to specially emphasize that the proposed techniques might be useful for investigations of the internal gettering process.[35,36] The presence of the gettering precipitates can be checked by OLALS with quasi-bulk photoexcitation. The parameters of impurity atmospheres around the gettering defects can be studied using the SLALS microscope in its basic mode. The denuded zone may be checked by OLALS with surface photoexcitation. A composition of the impurity atmospheres can be investigated using temperature dependencies of LALS intensity.

Investigations of different processes of gettering are possible by the LALS-based techniques too.[88] The procedures for these studies are analogous to those described above for the internal gettering process.

5. *Evaluation of processing effects on carrier lifetime.* The excess carrier lifetime and surface recombination velocity evaluation is also possible by means of OLALS. This technique might be useful e.g. for identifying device fabrication processes or chemicals which may lower excess carrier lifetime or increase surface generation and recombination. OLALS is very attractive for this purpose because it gives local values of the evaluated parameters—that enables the mapping—and because the measurements can be performed at any stage of device processing without destruction.

6. *Investigation of materials for photovoltaics.* Investigation of Si and different single- and multicrystalline semiconducting materials for solar cells including the non-equilibrium carrier lifetime measurements and the inspection of the grain boundaries passivation efficiency are possible applications of OLALS too.[74]

So, we can conclude that the techniques based on LALS might by very effective non-destructive tools for solving a wide range of problems of semiconducting materials and structures testing in microelectronics, optoelectronics and photovoltaics, and might be used both in laboratories and directly in the production cycle. We have mentioned only several of the most obvious potential applications of these techniques. We are sure, however, that they might find a great number of additional applications and be useful in many branches of semiconductor science and industry.

It should be emphasized also that the proposed instrument is a simple, inexpensive, and convenient to use, diagnostic device for express testing of materials and structures in the industry. In addition, a number of methods for automated in-line quality inspection and remote process monitoring might be developed on the basis of LALS.

This research has been financed in part by the Russian Foundation for Basic Research (grant No. 96–02–19540) and the sub-program "Perspective Technologies and Devices of Micro- and Nanoelectronics" of the Ministry of Science and Technologies of the Russian Federation (grant No. 02.04.3.2.40.Э.24). The authors express their appreciation for the financial support.

# FIGURE CAPTIONS

Fig. 1. Installation diagram for LALS with angular resolution: (1) $CO_2$ or CO-laser, (2,10) semi-transparent mirrors, (3,4) mirrors, (5) filters, (6) laser for photoexcitation (used in LALS with photoexcitation), (7) sample (can be placed in a cryostat or a furnace), (8) movable mirror, (9,11) photoreceivers, (12) computer.

Fig. 2. Typical light scattering diagrams for two CZ Si:B wafers; (*a*) as-grown, orientation with respect to the detection plane: (1) 0°, (2) 90°; (*b*) effect of internal gettering: (1) as-grown, (2) after gettering.

Fig. 3. Scattering diagrams for CZ Si at different wavelengths of the probe radiation $\lambda$: (1) 10.6 μm; (2) 5.4 μm; (3) 1.15 μm.

Fig. 4. Temperature dependencies of the light scattering intensity for different materials (material, estimated activation energies): (1) Si, $E_1 = 0.14$–$0.18$ eV; (2) Ge, $E_2 = 0.22$–$0.23$ eV, $E_3 = 0.11$–$0.13$ eV; (3) GaAs, $E_4 = 0.06$–$0.09$ eV; (4) InP, $E_5 = 0.02$–$0.05$ eV.

Fig. 5. Temperature dependencies of the light scattering intensity (*a*) and light scattering diagrams at different temperatures (*b*) for CZ Si:B single crystals: (*a*), defects: (1) cylindrical, $E_i - E_v = 120$–$160$ meV, (2) spherical, $E_i - E_v = 40$–$60$ meV; (*b*), sample temperature: (1) 300 K, (2) 110 K.

Fig. 6. Spatial potential relief calculated for the model of p-type LSDAs in n-type InP (1) and InP:Fe (2) crystals from the data of LALS; model parameters: donor concentration $N_D = 10^{16}$ cm$^{-3}$, maximum concentration of point defects forming the LSDA (assumed to be In$_P$, $E_i - E_v \approx 30$ meV) $N_{i\,max} = 2 \times 10^{17}$ cm$^{-3}$, iron concentration: (1) $N_{Fe} = 0$, (2) $N_{Fe} = 1.1 \times 10^{16}$ cm$^{-3}$, $N_i = N_{i\,max} \exp(-r^2/a^2)$; 300 K.

Fig. 7. (*a*) Light scattering diagrams measured without (1) and with (2),(3) band-to-band photoexcitation for mechanically (2) and chemical-mechanically (3) lapped wafers of Ge; $\lambda = 10.6$ μm, excitation data: $\lambda_{ex} = 1.06$ μm, pulse duration $\tau_p = 50$ ns, energy

$E_p = 0.3$ mJ, frequency $f_p = 1$ kHz, spot diameter $d_s = 8$ mm. (*b*) Light scattering intensity vs excitation power, mechanically polished sample.

Fig. 8. Light scattering diagrams for CZ Si:B measured under quasi-bulk photoexcitation before (1) and after (2) the internal gettering treatment; $\lambda = 10.6$ μm, $\lambda_{ex} = 1.06$ μm.

Fig. 9. The pictorial optical diagram of the SLALS microscope: (1) the probe IR-laser beam (routinely, CO or $CO_2$-laser), (2) the studied sample (can be placed in a cryostat or a furnace) in the front focal plane of the lens L1, (3) an aperture with diameter $D_1$ of the lens L1, (4) the lens L1, (5) an opaque screen or a mirror with a radius $b_1$ in the back focal plane of the lens L1, (6) an aperture with a radius $b_0$ in the back focal plane of the lens L1, (7) an aperture with a diameter $D_2$ of the lens L2, (8) the lens L2, (9) the scattered wave, (10) an IR photodetector in the back focal plane of the lens L2, (11) an exciting light beam (used in the OLALS mode).

Fig. 10. Pairs of SLALS (*a*),(*c*),(*e*),(*g*) and OLALS (*b*),(*d*),(*f*),(*h*) micrographs of the same regions of CZ Si:P wafers from the CCD matrix manufacturing cycle (1×1 mm); (*a*),(*b*): initial wafer; (*c*),(*d*): under $SiO_2$ layer (1200 Å thick); (*e*),(*f*): under $SiO_2$ and $Si_3N_4$ layers; (*g*),(*h*): CCD chip.

Fig. 11. Couples of SLALS (*a*),(*c*),(*e*),(*g*),(*i*),(*k*) and OLALS (*b*),(*d*),(*f*),(*h*),(*j*),(*l*) micrographs of the same regions of $Si_{1-x}Ge_x$ wafers (1×1 mm); (*a*)–(*d*): p-type CZ Si (100), 4 at. % of Ge; (*e*)–(*h*): p-type CZ Si (111), 4.7 at. % of Ge; (*i*)–(*l*): n-type CZ Si (111), 2.2 at. % of Ge; (*e*),(*f*),(*i*),(*j*) are close to the wafer centers, the rest are far from the centers.

Fig. 12. OLALS micrographs of multicrystalline silicon for solar cells (the optical beam induced absorption, 4×4 mm). The darker the image, the shorter the lifetime is. Grain boundaries are clearly seen in the micrographs.

Fig. 13. OLALS: dependencies of the detector signal for scattering (1),(2) and absorption (3) sub-modes on the power of the exciting laser beam in the crystal for chemical-

mechanically (1),(3) and mechanically (2) polished sides of a Si wafer; $\lambda = 10.6$ µm, $\lambda_{ex} = 0.63$ µm.

however. (Compare also with the dark-field heterodyne microscope for which a detector

signal $I_d \sim \sqrt{W} \left| \tilde{\varepsilon}(x,y) \right|$.[3,52])

[83] It is easy to show that if the impact of charged extended linear defects with dimensions $L_x$

and $L_y$ ($L_y \gg L_x$) to the carrier scattering prevails and the pulse relaxation time

$\tau \ll 1/\omega$ due to high defect concentration, then $\tau \sim \left(1/L_{\mathrm{D}} L_y\right) \sim n^{1/2}$ and

$I \sim \left[\mathrm{Im}(\delta\varepsilon)\right]^2 \sim \left[\mathrm{Im}\,\varepsilon\right]^2 \sim \omega_{\mathrm{p}}^4 \tau^2 \sim n^3 \sim W_{ex}^3$; $\omega_{\mathrm{p}}$ is the plasma frequency; the Debye length

$L_D$ is assumed to satisfy the condition: $L_y \gg L_{\mathrm{D}} \gg L_x$; the linear recombination is implied.

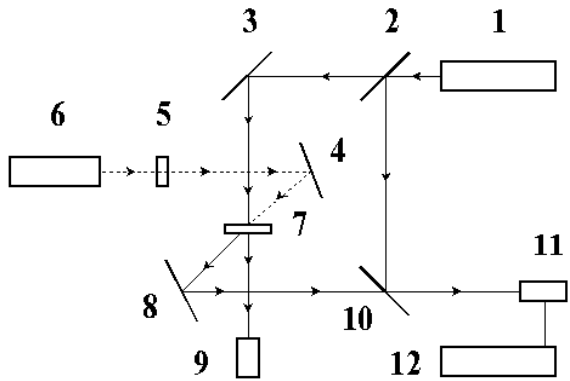

Fig. 1. Installation diagram for LALS with angular resolution: (1) $CO_2$ or CO-laser, (2,10) semi-transparent mirrors, (3,4) mirrors, (5) filters, (6) laser for photoexcitation (used in LALS with photoexcitation), (7) sample (can be placed in a cryostat or a furnace), (8) movable mirror, (9,11) photoreceivers, (12) computer.

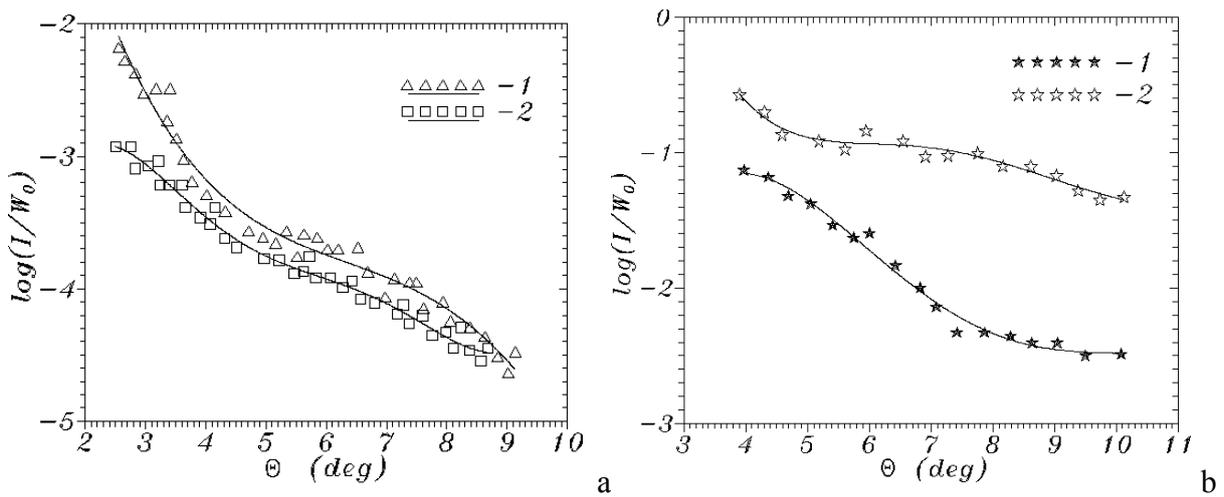

Fig. 2. Typical light scattering diagrams for two CZ Si:B wafers; (*a*) as-grown, orientation with respect to the detection plane: (1) 0°, (2) 90°; (*b*) effect of internal gettering: (1) as-grown, (2) after gettering.

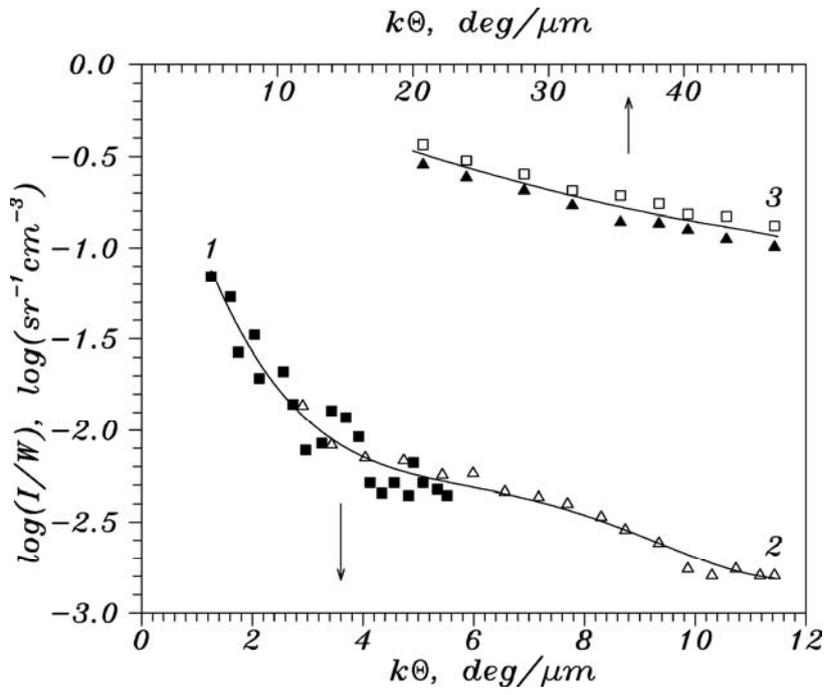

Fig. 3. Scattering diagrams for CZ Si at different wavelengths of the probe radiation $\lambda$: (1) 10.6 µm; (2) 5.4 µm; (3) 1.15 µm.

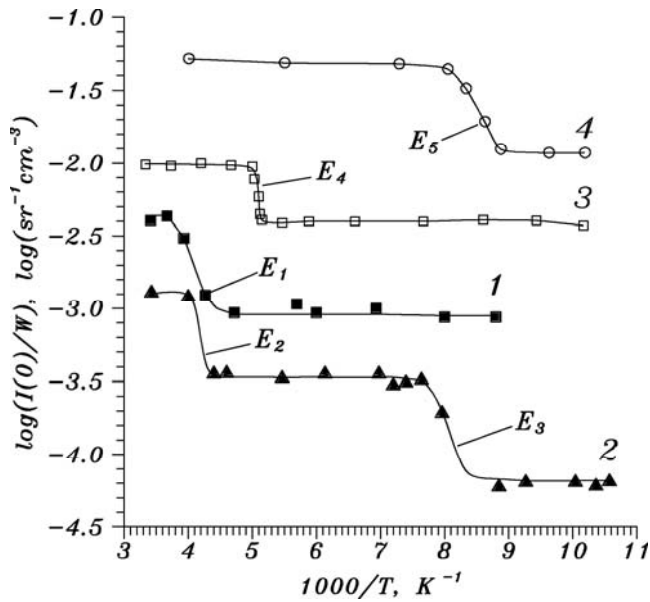

Fig. 4. Temperature dependencies of the light scattering intensity for different materials (material, estimated activation energies): (1) Si, $E_1 = 0.14$–$0.18$ eV; (2) Ge, $E_2 = 0.22$–$0.23$ eV, $E_3 = 0.11$–$0.13$ eV; (3) GaAs, $E_4 = 0.06$–$0.09$ eV; (4) InP, $E_5 = 0.02$–$0.05$ eV.

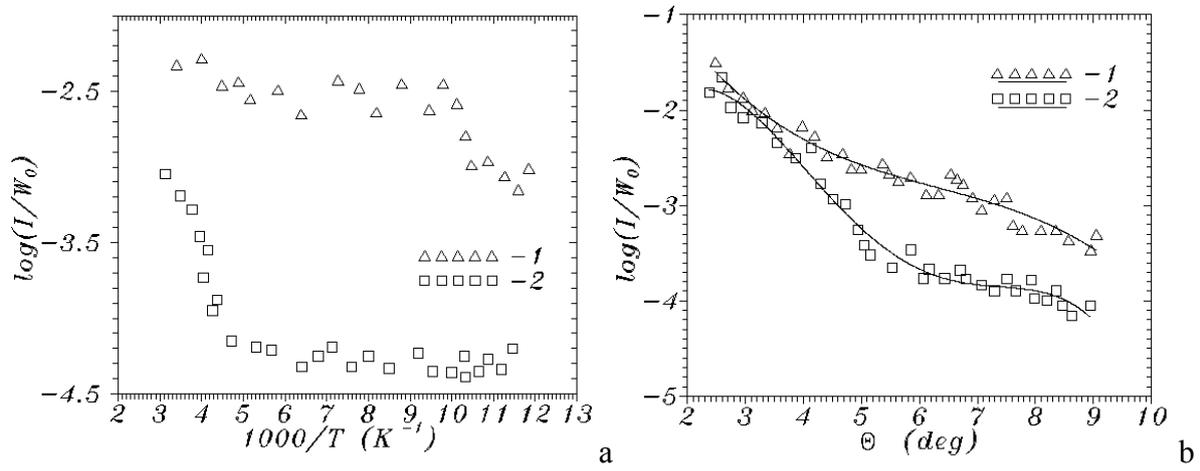

Fig. 5. Temperature dependencies of the light scattering intensity (*a*) and light scattering diagrams at different temperatures (*b*) for CZ Si:B single crystals: (*a*), defects: (1) cylindrical, $E_i - E_v$ = 120–160 meV, (2) spherical, $E_i - E_v$ = 40–60 meV; (*b*), sample temperature: (1) 300 K, (2) 110 K.

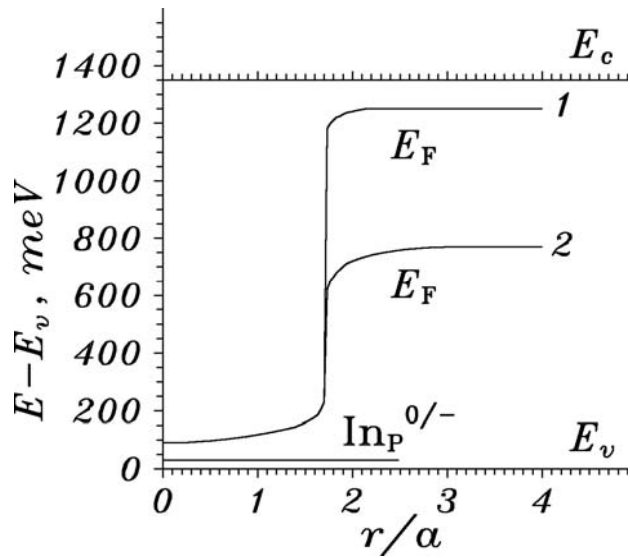

Fig. 6. Spatial potential relief calculated for the model of p-type LSDAs in n-type InP (1) and InP:Fe (2) crystals from the data of LALS; model parameters: donor concentration $N_D = 10^{16}$ cm$^{-3}$, maximum concentration of point defects forming the LSDA (assumed to be InP, $E_i - E_v \approx 30$ meV) $N_{i\,max} = 2 \times 10^{17}$ cm$^{-3}$, iron concentration: (1) $N_{Fe} = 0$, (2) $N_{Fe} = 1.1 \times 10^{16}$ cm$^{-3}$, $N_i = N_{i\,max} \exp(-r^2/a^2)$; 300 K.

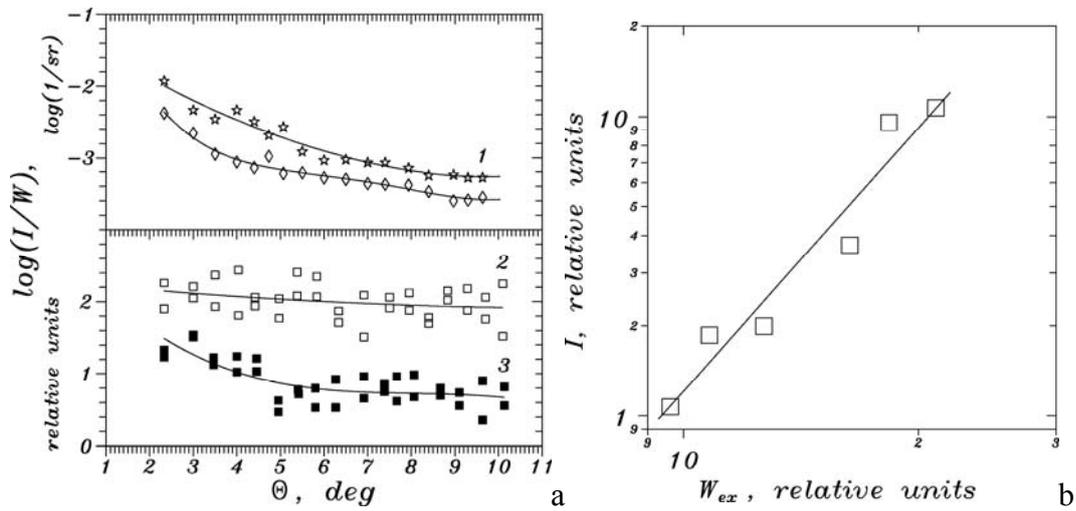

Fig. 7. (*a*) Light scattering diagrams measured without (1) and with (2),(3) band-to-band photoexcitation for mechanically (2) and chemical-mechanically (3) lapped wafers of Ge; $\lambda = 10.6$ μm, excitation data: $\lambda_{ex} = 1.06$ μm, pulse duration $\tau_p = 50$ ns, energy $E_p = 0.3$ mJ, frequency $f_p = 1$ kHz, spot diameter $d_s = 8$ mm. (*b*) Light scattering intensity vs excitation power, mechanically polished sample.

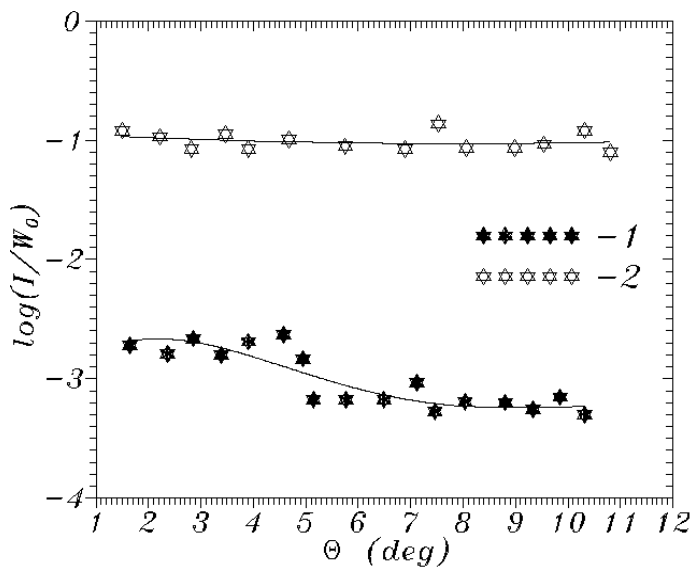

Fig. 8. Light scattering diagrams for CZ Si:B measured under quasi-bulk photoexcitation before (1) and after (2) the internal gettering treatment; $\lambda = 10.6$ μm, $\lambda_{ex} = 1.06$ μm.

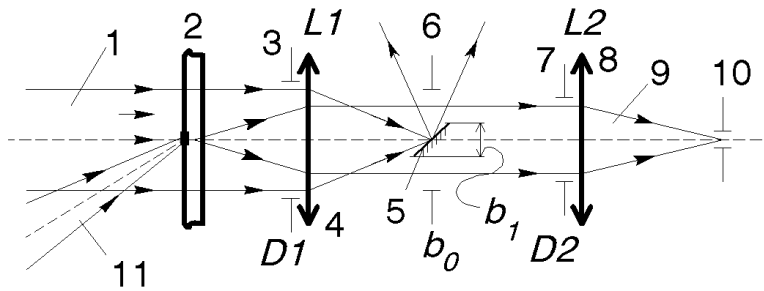

Fig. 9. The pictorial optical diagram of the SLALS microscope: (1) the probe IR-laser beam (routinely, CO or $CO_2$-laser), (2) the studied sample (can be placed in a cryostat or a furnace) in the front focal plane of the lens L1, (3) an aperture with diameter $D_1$ of the lens L1, (4) the lens L1, (5) an opaque screen or a mirror with a radius $b_1$ in the back focal plane of the lens L1, (6) an aperture with a radius $b_0$ in the back focal plane of the lens L1, (7) an aperture with a diameter $D_2$ of the lens L2, (8) the lens L2, (9) the scattered wave, (10) an IR photodetector in the back focal plane of the lens L2, (11) an exciting light beam (used in the OLALS mode).

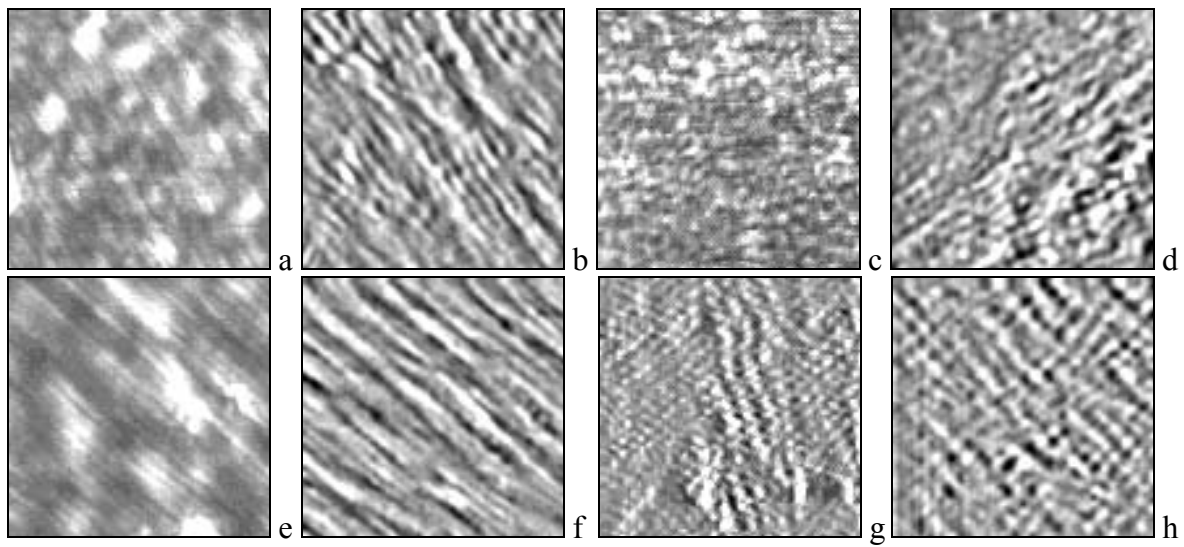

Fig. 10. Pairs of SLALS ($a$),($c$),($e$),($g$) and OLALS ($b$),($d$),($f$),($h$) micrographs of the same regions of CZ Si:P wafers from the CCD matrix manufacturing cycle (1×1 mm); ($a$),($b$): initial wafer; ($c$),($d$): under $SiO_2$ layer (1200 Å thick); ($e$),($f$): under $SiO_2$ and $Si_3N_4$ layers; ($g$),($h$): CCD chip.

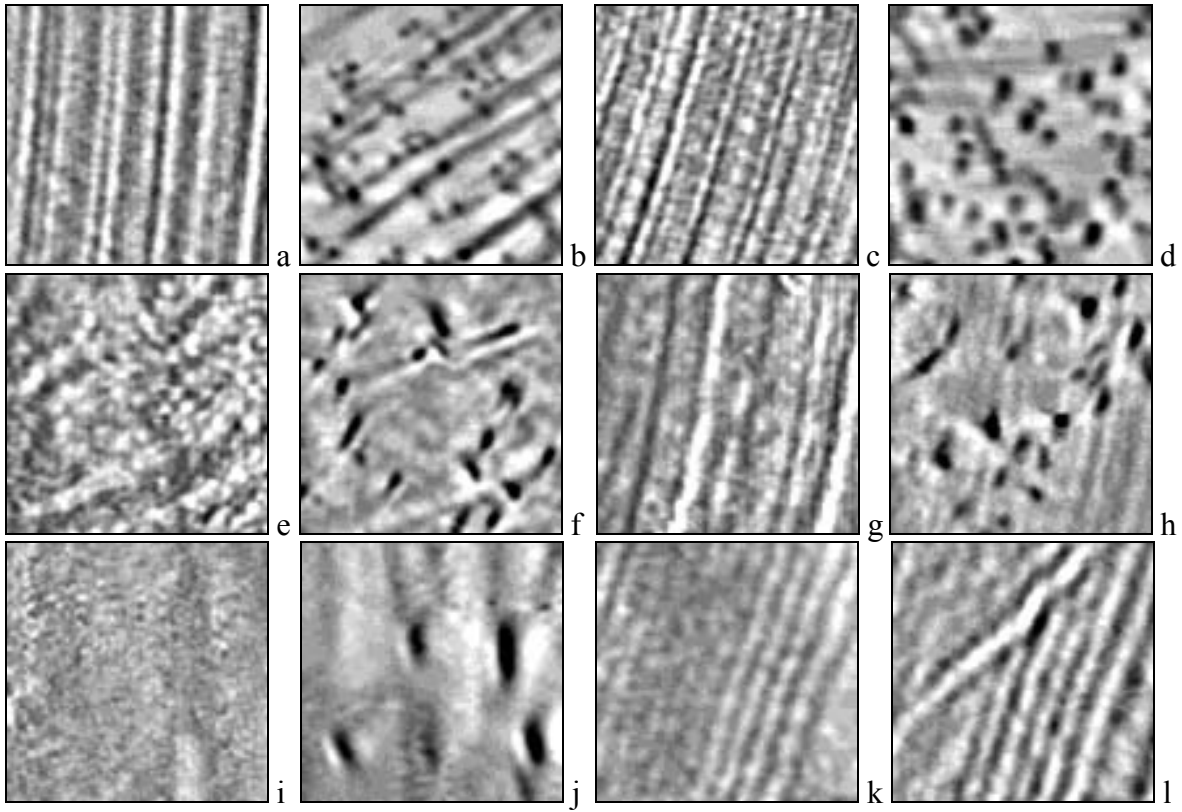

Fig. 11. Couples of SLALS (*a*),(*c*),(*e*),(*g*),(*i*),(*k*) and OLALS (*b*),(*d*),(*f*),(*h*),(*j*),(*l*) micrographs of the same regions of $Si_{1-x}Ge_x$ wafers ($1 \times 1$ mm); (*a*)–(*d*): p-type CZ Si (100), 4 at. % of Ge; (*e*)–(*h*): p-type CZ Si (111), 4.7 at. % of Ge; (*i*)–(*l*): n-type CZ Si (111), 2.2 at. % of Ge; (*e*),(*f*),(*i*),(*j*) are close to the wafer centers, the rest are far from the centers.

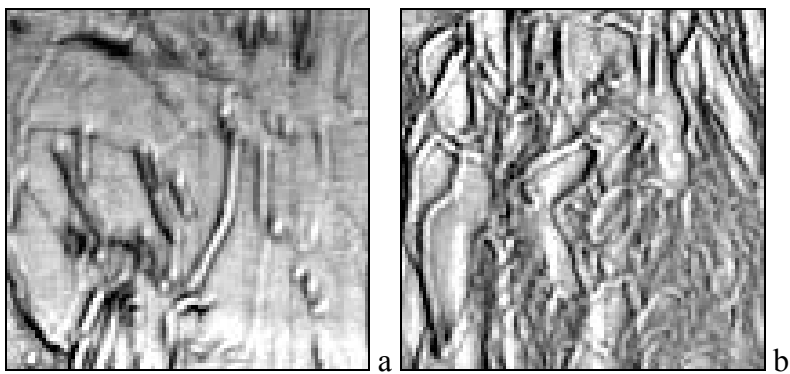

Fig. 12. OLALS micrographs of multicrystalline silicon for solar cells (the optical beam induced absorption, $4 \times 4$ mm). The darker the image, the shorter the lifetime is. Grain boundaries are clearly seen in the micrographs.

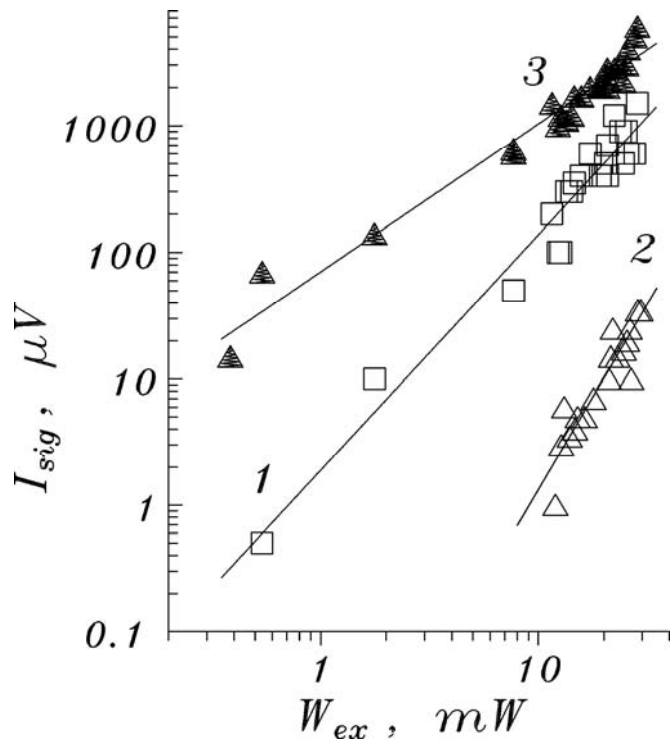

Fig. 13. OLALS: dependencies of the detector signal for scattering (1),(2) and absorption (3) sub-modes on the power of the exciting laser beam in the crystal for chemical-mechanically (1),(3) and mechanically (2) polished sides of a Si wafer; $\lambda = 10.6$ μm, $\lambda_{ex} = 0.63$ μm.